\documentclass[aps,preprint,nofootinbib,superscriptaddress]{revtex4}

\usepackage{amssymb}

\usepackage{epsfig}
\usepackage[bookmarksnumbered,bookmarksopen,colorlinks,citecolor=blue,linkcolor=blue]{hyperref}

\begin{document}

\title{ New neutron-rich isotope production in $^{154}$Sm+$^{160}$Gd }

\author{Ning Wang}
\email{wangning@gxnu.edu.cn}\affiliation{ Department of Physics,
Guangxi Normal University, Guilin 541004, People's Republic of
China }

\author{Lu Guo}
\email{luguo@ucas.ac.cn} \affiliation{ School of Physics, University of Chinese Academy of Sciences, Beijing 100049, People's Republic of China}
\affiliation{ State Key Laboratory of Theoretical Physics, Institute of Theoretical Physics, Chinese Academy of Sciences, Beijing 100190, People's Republic of China}

\begin{abstract}

Deep inelastic scattering in $^{154}$Sm+$^{160}$Gd at energies above the Bass barrier is for the first time investigated with two different microscopic dynamics approaches: improved quantum molecular dynamics (ImQMD) model and time dependent Hartree-Fock (TDHF) theory. No fusion is observed from both models. The capture pocket disappears for this reaction due to  strong Coulomb repulsion and the contact time of the di-nuclear system formed in head-on collisions is about 700 fm/c at an incident energy of 440 MeV. The isotope distribution of fragments in the deep inelastic scattering process is predicted with the simulations of the latest ImQMD-v2.2 model together with a statistical code (GEMINI) for describing the secondary decay of fragments. More than 40 extremely neutron-rich unmeasured nuclei with $58 \le Z\le 76$ are observed and the production cross sections are at the order of ${\rm \mu b}$ to mb. The multi-nucleon transfer reaction of Sm+Gd could be an alternative way to synthesize new neutron-rich lanthanides which are difficult to be produced with traditional fusion reactions or fission of actinides.

\end{abstract}
\maketitle

\begin{center}
\textbf{I. INTRODUCTION}
\end{center}

The heavy-ion reaction at energies around the Coulomb barrier is an
important way not only for the study of the nuclear structures, but
also for the synthesis of unstable or even exotic (neutron-rich,
neutron-deficient, superheavy) nuclei for which no experimental data
exist \cite{OPb2,OSm,Timm,Zhang10,Hof00,Ogan10,Sob,Gup05}. For light
and intermediate fusion systems, the fusion process is usually
described by the penetration of the fusion barriers. The fusion
(capture) cross sections can be accurately predicted by using the
fusion coupled channel calculations or empirical barrier
distribution approaches
\cite{Wong73,Hag99,liumin06,Wang09,Gup92,Wangbin}. For fusion
systems leading to the synthesis of super-heavy elements, the
quasi-fission and fusion-fission process significantly complicates
the description of fusion process. The very shallow capture pockets
in such kind of reaction systems may cause some difficulties in the
applications of the barrier-penetration approaches. Although
macroscopic dynamics models  \cite{Shen02,Zagr05,
Adam97,Diaz01,Nan12} met with some success for describing the
residual evaporation cross sections of measured super-heavy systems,
the uncertainty of the predicted fusion probability from these
different models  for unmeasured systems is still large due to the
uncertainty of model parameters \cite{Nas11,Bao15} and ambiguity of
reaction mechanism. For example, with the fusion-by-diffusion model,
Choudhury and Gupta \cite{Chou14} investigated symmetric heavy-ion
reaction of $^{154}$Sm+$^{154}$Sm and obtained measurable
evaporation residue cross sections ($\sim$ 0.6 pb). However, Cap et
al. \cite{Cap14} investigated the same reaction and found the cross
sections are extremely small (about 10$^{-13}$ pb ) and probably
never reachable. The contradictory predictions imply some key model
parameters such as the injection point distance and the dynamical
nucleus-nucleus potential are far from clear for this reaction. It
is therefore necessary to investigate the dynamics process and
fusion probability in this kind of reactions with self-consistent
microscopic dynamics models.

In addition to the formation of superheavy nuclei, the synthesis of
extremely neutron-rich heavy nuclides through multi-fragmentation,
deep inelastic scattering and quasi-fission are of exceptional
importance to advance our understanding of nuclear structure at the
extreme isospin limit of the nuclear landscape
\cite{Shen87,Ober14,Zag11,Sou03,Heinz14}. Neutron-rich lanthanides,
such as $^{182}_{~70}$Yb$_{112}$ with "false magic numbers", are of
importance for understanding the strength of spin-orbit interaction
which influences the positions of the island of stability for
super-heavy nuclei. Unfortunately, if one glances at the chart of
nuclides (see the positions of known nuclei in AME2012
\cite{Audi12}), one notes that the number of observed neutron-rich
nuclides is very limited at mass region $A>160$, due to that neither
traditional fusion reactions with stable beams nor fission of
actinides easily produce new neutron-rich heavy nuclei in this
region. Recently, some heavy neutron-rich nuclei with $70 \le Z \le
79$  were produced in projectile fragmentation of $^{197}$Au primary
beams bombarding on thick $^{9}$Be target at GSI \cite{Shu03}. In
addition to the fragmentation of heavy nuclei, multi-nucleon
transfer process might be helpful to produce  neutron-rich heavy
nuclei \cite{Noren74,Zhang98,Zag08,Vol78}. Zychor et al. have
performed a systematic study on the productions of Hafnium and
Lutetium isotopes with the reactions induced on a thick tungsten
target by $^{40}$Ar, $^{84}$Kr, and $^{136}$Xe, respectively. The
study indicated that the absolute production cross sections of
neutron-rich heavy isotopes increase with increasing projectile mass
\cite{Zych84}. It is necessary and important to study the
multi-nucleon transfer between two nuclei in the rare-earth region
for producing new neutron-rich lanthanides, considering the fission
barriers of lanthanides are relatively high to prevent fission of
heavy fragments in the secondary decay process. The investigation of
deep inelastic scattering in $^{154}$Sm+$^{160}$Gd at energies above
the Coulomb barrier is therefore interesting, not only for the study
of the production probability of super-heavy nuclei, but also for
the synthesis of unmeasured lanthanides.

To understand the dynamical process in fusion and deep inelastic
scattering reactions, some microscopical dynamics models, such as
the time-dependent Hartree-Fock (TDHF)
\cite{Naka05,Maru06,Guo07,Sim12,Umar12,Sim14,Stev16} and some
different extended versions of quantum molecular dynamics (QMD)
model \cite{QMD} including IQMD \cite{IQMD93,IQMD99}, CoMD
\cite{constrain,Maru02,Sou14}, ImQMD
\cite{ImQMD2002,ImQMD2004,ImQMD2010,ImQMD2014}, EQMD
\cite{Maru96,Ma14}, etc. have been developed. TDHF theory has many
successful applications in the description of nuclear large
amplitude collective motions, for example, heavy-ion collisions,
giant resonance, fission dynamics, and nuclear molecular resonance;
for a recent review see Ref. \cite{Sim14}. TDHF in a nuclear context
means a time-dependent mean-field theory derived from an effective
energy functional. The most widely used is the Skyrme energy density
functional (EDF) which leads to an accurate description of selected
static properties in nuclei. Static and dynamical mean-field
theories, by considering directly single-particle degrees of freedom
interacting, was a major breakthrough in nuclear physics to describe
static and dynamical nuclear properties \cite{Lac15}. Comparing with
the semi-classic molecular dynamics simulations, the TDHF
calculations can describe better the structure effects of nuclear
system such as the shell effects and nuclear shapes in heavy-ion
reaction at low incident energies.

On the other hand, at the very early stage of the application of
TDHF, it was already realized that the independent particle picture
used in the mean-field theory leads to severe limitations
\cite{Lac15}. It is known the one-body microscopic dynamics models
based on the mean-field theory are difficult to describe a
multi-fragmentation process, due to the fact that the correlations
treated in the one-body approach are not able to describe the large
fluctuations \cite{constrain}. This difficulty can be solved by
adopting more suitable treatments of the $N$-body problem like
molecular dynamics. In the improved quantum molecular dynamics
(ImQMD) model, the standard Skyrme force with the omission of
spin-orbit term is adopted for describing not only the bulk
properties but also the surface properties of nuclei.
Simultaneously, the Fermi constraint is used to maintain the
fermionic feature of the nuclear system. In the Fermi constraint
which was previously proposed by Papa et al. in the CoMD model
\cite{constrain} and improved very recently in Refs.
\cite{Wang15,Wang16}, the phase space occupation probability $\bar
f_{i}$ of the $i$-th particle is checked during the propagation of
nucleons. If $\bar f_{i}>1$, i.e. violation of the Pauli principle,
the momentum of the particle $i$ is randomly changed by a series of
two-body "elastic scattering" and "inelastic scattering" between
this particle and its neighboring particles, together with Pauli
blocking condition being checked after the momentum re-distribution.
In other words, both the self-consistently generated mean-field and
the momentum re-distribution in the Fermi constraint which
introduces additional fluctuations and two-body dissipation affect
the movements of nucleons in the simulations. The ImQMD model allows
to investigate the formation of fragments during a heavy-ion
collision in a consistent $N$-body treatment, through event-by-event
simulations, with which the charge and isotope distributions of
fragments can be obtained.

Considering the advantage of the TDHF theory in the description of
nuclear structure effects and that of the ImQMD model in the
description of fluctuations and fragment formation, it is therefore
necessary to investigate the same reaction system with these two
different microscopic dynamics models, for exploring the dynamical
mechanism and improving the reliability of model predictions for
unmeasured reaction systems.

The structure of this paper is as follows: In sec. II, the
frameworks of TDHF and ImQMD will be introduced. In sec. III, the
deep inelastic scattering process of $^{154}$Sm+$^{160}$Gd at an
incident energy of $E_{\rm c.m.}=440 $ MeV will be investigated with
the two models. In Sec. IV, the isotope distribution, angular
distribution and production cross sections of some neutron-rich
nuclei with unmeasured masses will be studied with the ImQMD model.
Finally a brief summary is given in Sec. V.

\begin{center}
\textbf{ II. Theoretical Frameworks }\\
\end{center}

In the TDHF theory, the complicated many-body problem is replaced by
an independent particle problem, i.e., the many-body wave functions are approximated as the anti-symmetrized
independent particle states to assure an exact
treatment of Pauli principle during time evolution. In the nuclear context, the basic ingredient of TDHF is
the energy functional composed by the various one-body densities. Here, we adopt the full Skyrme EDF with the parameter set SLy5 \cite{SLy5}. The Skyrme parameters have been fitted with the ground state properties of the selected nuclei. For the heavy-ion collisions, there is no adjustable free parameters in TDHF. The dynamical evolution of the mean-field is expressed by TDHF equation
\begin{equation}
i \hbar \frac{d \hat{\rho}}{dt} =[ \hat{h}[\hat{\rho}],\hat{\rho}],
\end{equation}
with the single-particle Hamiltonian $h[\hat\rho]$ and the one-body
density $\hat{\rho}$. Taking the nuclear ground state as an initial
state of the dynamical evolution, TDHF time evolution is determined
by the dynamical unitary propagator. Earlier TDHF calculations
imposed the various approximations on the effective interaction and
geometric symmetry. The development of computational power allows a
fully three-dimensional (3D) TDHF calculation with the modern
effective interaction and without symmetry restrictions, which
significantly improves the physical scenario in heavy-ion collisions
\cite{Guolu14}. In this work, the set of nonlinear TDHF equation is
solved on a three-dimensional Cartesian coordinate-space without any
symmetry restrictions. We use the fast Fourier transformation method
to calculate the derivatives. The conservation of total energy and
particle number is assured during the time evolution by choosing the
parameters of grid spacing as 1 fm and time step $\Delta t=0.2$
fm/c.

In the ImQMD simulations, each
nucleon is represented by a coherent state of a Gaussian wave
packet
\begin{equation}
\phi _{i}(\mathbf{r})=\frac{1}{(2\pi \sigma _{r}^{2})^{3/4}}\exp \left [-\frac{(%
\mathbf{r}-\mathbf{r}_i)^{2}}{4\sigma _{r}^{2}} +\frac{i}{\hbar} \mathbf{r} \cdot \mathbf{p}_i \right ],
\end{equation}
where $\mathbf{r}_{i}$ and $\mathbf{p}_{i}$ are the centers of the $i$-th
wave packet in the coordinate and momentum space, respectively.
$\sigma _{r}$ represents the spatial spread of the wave packet.
The total $N$-body wave function is assumed to be the direct product
of these coherent states. The anti-symmetrization effects are additionally simulated by introducing the Fermi constraint mentioned previously (In the traditional QMD calculations, the Pauli potential \cite{Pei92} or momentum-dependent two-body repulsion \cite{Wilets,Dorso87} and the collision term \cite{Rein96} are usually used to simulate the effects). Through a Wigner transformation, the one-body  phase space distribution function and the density distribution function $\rho$ of a system
\begin{equation} \label{1}
\rho(\mathbf{r})=\sum_i{\frac{1}{(2\pi \sigma_r^2)^{3/2}}\exp
\left [-\frac{(\mathbf{r}-\mathbf{r}_i)^2}{2\sigma_r^2} \right ]},
\end{equation}
are obtained. The propagation of nucleons is governed by the self-consistently generated mean-field,
\begin{equation} \label{2}
\mathbf{\dot{r}}_i=\frac{\partial H}{\partial \mathbf{p}_i}, \; \;
\mathbf{\dot{p}}_i=-\frac{\partial H}{\partial \mathbf{r}_i},
\end{equation}
and the momentum re-distribution in the Fermi constraint. Euler algorithm is adopted to compute new positions and momenta at time $t+ \Delta t$. The time step in the ImQMD calculations is set as $\Delta t=1$ fm/c. The Hamiltonian $H$ consists of the kinetic energy and the effective
interaction potential energy which is based on the Skyrme EDF by neglecting the spin-orbit term. The model parameter set IQ3a \cite{ImQMD2014} is adopted in present ImQMD calculations. With an incompressibility coefficient of about 225 MeV for symmetric nuclear matter, IQ3a is suitable for the description of heavy-ion collisions at intermediate and low energies such as fusion reactions at energies around the Coulomb barrier \cite{ImQMD2014,Wang14a} and multi-fragmentation at Fermi energies \cite{Wang16}.

\begin{center}
\textbf{IV. Dynamical Scattering Process in $^{154}$Sm+$^{160}$Gd}
\end{center}

\begin{figure}
\includegraphics[angle=0,width=0.9 \textwidth]{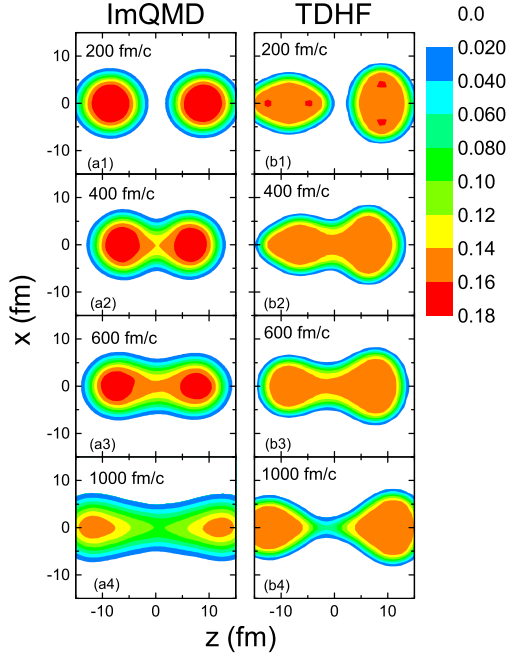}
\caption{(Color online) Time evolution of the density distribution in head-on collision of $^{154}$Sm+$^{160}$Gd at center-of-mass incident energy $E_{\rm c.m.}=440$ MeV. The left sub-figures denote the results of the ImQMD simulations and the right ones denote those of TDHF.}
\end{figure}

For the nearly-symmetric reaction $^{154}$Sm+$^{160}$Gd, the height of Coulomb barrier from Bass potential \cite{Bass80} is 393 MeV and the predicted Q-value for complete fusion is $Q=-410$ MeV according to the WS4 calculations \cite{Wang14}. Here, we first investigate the head-on collisions of $^{154}$Sm+$^{160}$Gd at $E_{\rm c.m.}=440$ MeV which is higher than the Bass barrier by 47 MeV. If the compound nuclei ($A=314$, $Z=126$) can be formed in such a reaction, the excitation energy of the compound nuclei at $E_{\rm c.m.}=440$ MeV  is only about 30 MeV and the residual nuclei might survive against fission during the de-excitation process.

\begin{figure}
\includegraphics[angle=0,width=0.75 \textwidth]{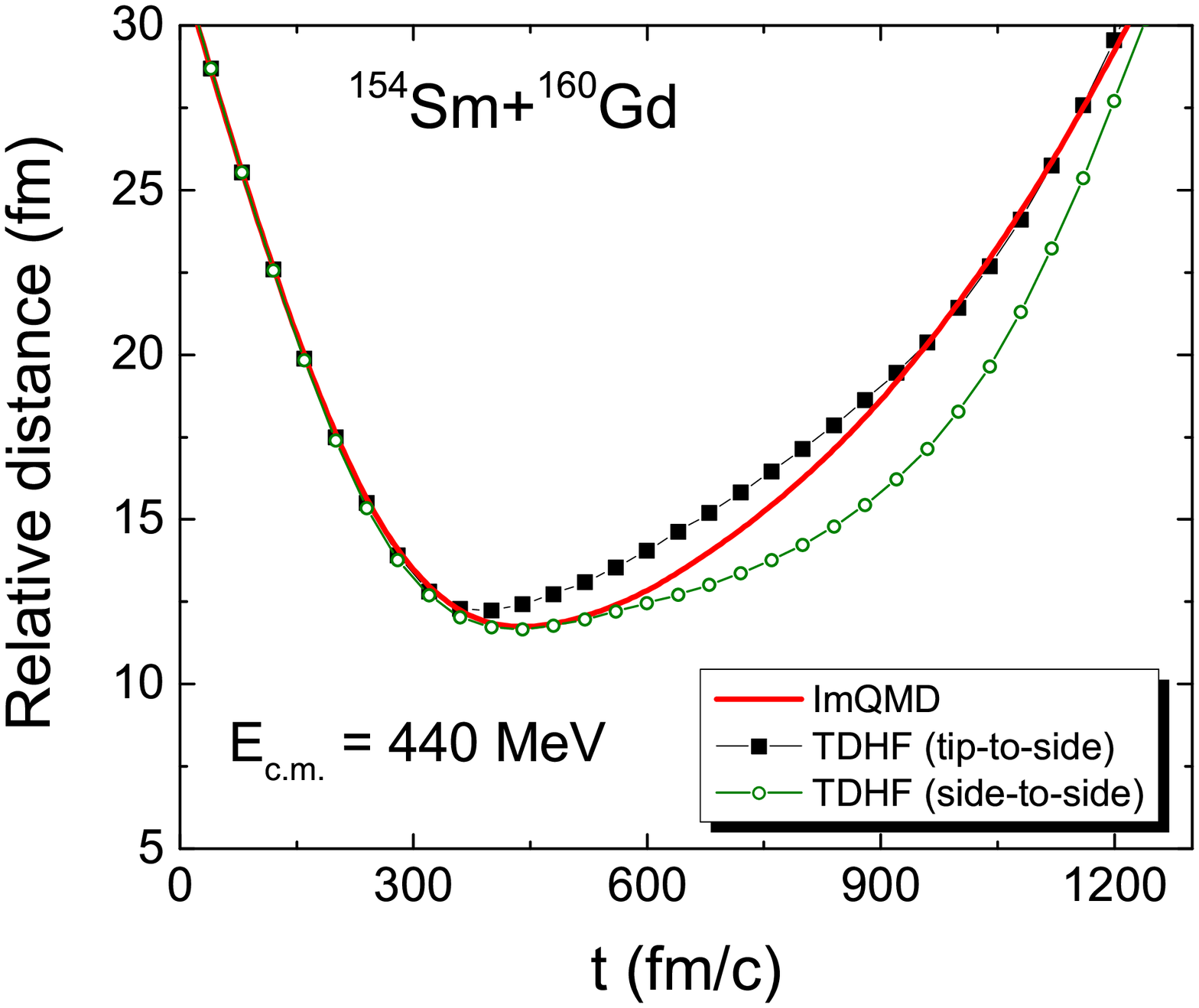}
\caption{(Color online) Time evolution of the relative distance between two nuclei in $^{154}$Sm+$^{160}$Gd. The solid curve denote the results of ImQMD. The squares and the circles denote the results of TDHF at tip-to-side and side-to-side collisions, respectively.}
\end{figure}

Fig. 1 shows the time evolution of the density distribution in head-on collision of $^{154}$Sm+$^{160}$Gd. The left panels denote the results of the ImQMD simulations. Because the shell effects of reaction partners are not considered self-consistently in the ImQMD simulations, the sampled initial nuclei are spherical in shape. In the TDHF calculations, the deformation effects of initial nuclei can be remarkably well described. For the orientation of the deformed projectile and target nuclei, we set the initial orientation as tip-to-side configuration. We note that the compound nuclei are not formed in the two different dynamics simulations, even the incident energy is obviously higher than the Bass barrier. At $t=1000$ fm/c, the neck of the di-nuclear system (DNS) becomes narrow and the system tend to split up. The contact-times of the DNS in this reaction are about 700 fm/c, which is much shorter than the typical contact-times of quasi-fission (usually greater than 1500 fm/c but much shorter than typical fusion-fission times) \cite{Shen87,Ober14}. Here, we would like to emphasize that the density distributions from the ImQMD simulations in Fig. 1 represent the average value over a large number simulation events. Simultaneously, we investigate the relative motion of $^{154}$Sm+$^{160}$Gd. Fig. 2 shows the time evolution of the relative distance between two nuclei. The squares and the circles denote the results of TDHF at different orientations for the deformed reaction partners, which are comparable with those of ImQMD (solid curve).  From Fig. 2, one sees that the results of TDHF are slightly different from those of ImQMD at touching configuration, which is due to the deformation effects of reaction partners in the TDHF simulations. From the time evolution of the relative distance at different orientations with the TDHF theory, we find that the results of side-to-side collision are slightly lower than those of tip-to-side collision by about $1 \sim 3$ fm at touching configuration, which indicates that the orientations of the deformed nuclei affect the time evolution of neck and the relative distance. Gupta et al. \cite{Gupta04} previously investigated the influence of the orientations of deformed nuclei on nuclear proximity potential and found that the resulting configuration is more compact at the side-to-side configuration, which is consistent with the TDHF calculations. The time evolution of relative motion between two nuclei indicates that deep inelastic scattering process plays a dominant role in the head-on collisions of $^{154}$Sm+$^{160}$Gd at $E_{\rm c.m.}=440$ MeV.   Due to the fluctuations and two-body dissipation \cite{Tian10,Koh80,Car86}, more phenomena are obtained in the ImQMD simulations, which is different from the TDHF picture. For example, we note that there is a tiny part of simulation events ($0.6\%$ at $b=3$ fm and $0.02\%$ at $b=5$ fm), in which the neck of the DNS still remains at $t=2000$ fm/c, although the DNS is strongly deformed and tend to split up. In addition, there is about $1\%$ of simulation events at central collisions in which the ternary breakup rather than traditional binary scattering is observed in the ImQMD simulations.

To understand the behavior of deep inelastic scattering at energies above the Bass barrier, it is necessary to investigate the nucleus-nucleus potential between these two nuclei. In heavy-ion fusion reactions, some static nucleus-nucleus potential are successfully proposed for describing the fusion barrier, such as the Bass potential \cite{Bass80} and the Woods-Saxon (WS) parametrization of the nuclear potential given by Broglia and Winther from a knowledge of the densities of the colliding nuclei and an effective two-body force \cite{BW91,Phoo13}. In addition, the Skyrme EDF together with extended Thomas-Fermi (ETF2) approach is also frequently used to investigate the Coulomb barrier based on the sudden approximation for the densities of the reaction partners \cite{liumin06,Wang09,Gup92}. Considering the uncertainty of these static/empirical potential at short distances, it is of importance to investigate the dynamical nucleus-nucleus potential. According to the energy conservation, we have \cite{ImQMD2010}
\begin{equation}
E_{\rm c.m.}=T+V+E^*+T_{\rm oth},
\end{equation}
where $E_{\rm c.m.}$ is the incident center-of-mass energy, $T$ is
the relative motion kinetic energy of two colliding nuclei, $E^*$ is the excitation energy, and $T_{\rm oth}$ is other
collective kinetic energy, such as vibrational energy of neck and rotational energy. Before the
neck of DNS being well formed in head-on collisions,
$E^*$ and $T_{\rm oth}$ could be negligible, the nucleus-nucleus potential is approximately
expressed as
\begin{equation}
V(R)\simeq E_{\rm c.m.}-T(R).
\end{equation}
At the closest distance $R_{\rm min}$, i.e. the smallest value in Fig. 2, $T(R_{\rm min})=0$. Therefore, the nucleus-nucleus potential at $R_{\rm min}$ can be roughly estimated by the corresponding incident energy, $V(R_{\rm min})\simeq E_{\rm c.m.}$, in the elastic and inelastic scattering collisions.

\begin{figure}
\includegraphics[angle=0,width=0.75\textwidth]{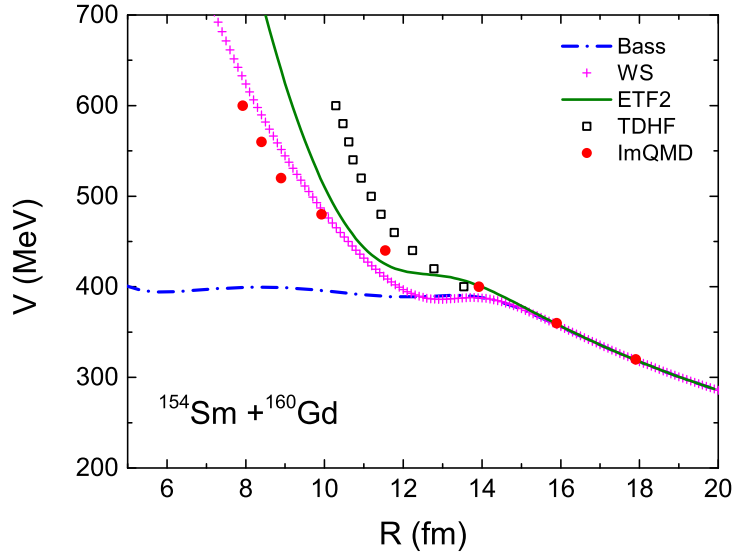}
\caption{(Color online) Nucleus-nucleus potential of $^{154}$Sm+$^{160}$Gd as a function of center-to-center distance between two nuclei. The circles denote the upper limit of the dynamical potential from ImQMD calculations and the squares denote those of TDHF at tip-to-side collisions.}
\end{figure}

In Fig. 3, we show the static nucleus-nucleus potential of $^{154}$Sm+$^{160}$Gd. The dot-dashed curve, the crosses and the solid curve denote the Bass potential, the Woods-Saxon potential of Broglia and Winther and the potential based on ETF2 approach, respectively. At the regions where two nuclei begin to touch each other ($R<14$ fm), Bass potential is flat, whereas the other two potentials suggest a strong repulsion between the reaction partners. The circles and squares denote the upper limit of the dynamical potential according to Eq.(6) together with calculated closest distance $R_{\rm min}$ from ImQMD and TDHF, respectively. Both the calculations of these static models and the results of ImQMD and TDHF indicate the capture pocket of this reaction system disappears in generally, which explains the deep inelastic scattering being the dominant process in head-on collisions of $^{154}$Sm+$^{160}$Gd at $E_{\rm c.m.}=440$ MeV. Even at much higher energies such as $E_{\rm c.m.}=600$ MeV, the fusion process is not observed with the two dynamics models. We also note that the contact time of the di-nuclear system (DNS) in central collisions is energy dependent. The contact time of the DNS increases with the incident energy in general.

 \begin{center}
\textbf{IV. Production of neutron-rich isotopes in $^{154}$Sm+$^{160}$Gd}
\end{center}

\begin{figure}
\includegraphics[angle=0,width=0.9\textwidth]{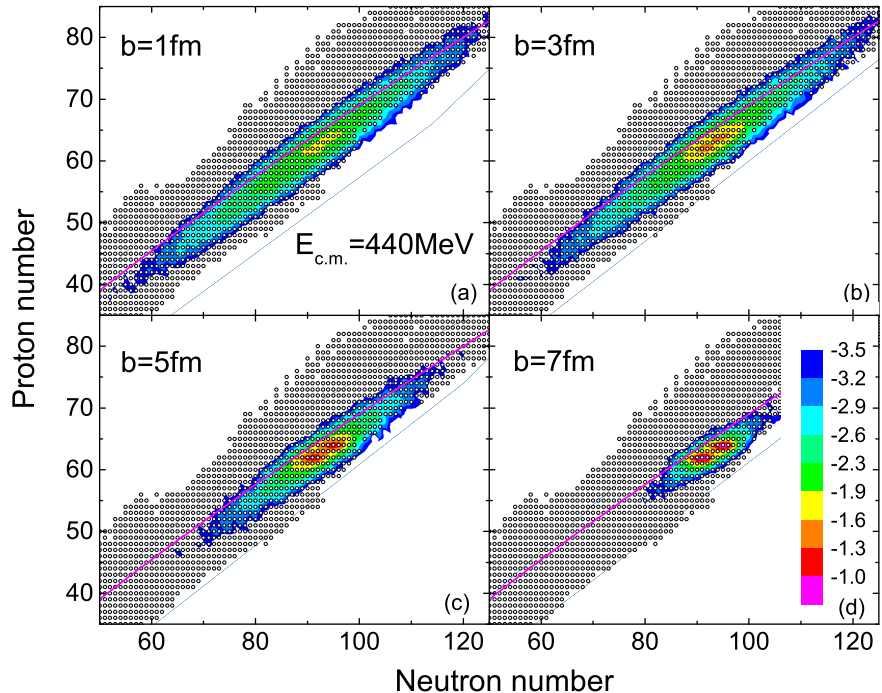}
\caption{(Color online) Isotope distribution of primary fragments in $^{154}$Sm+$^{160}$Gd at an incident energy of $E_{\rm c.m.}=440$ MeV. The circles denote the positions of known masses in AME2012.  }
\end{figure}

Although it is almost impossible to produce super-heavy nuclei in $^{154}$Sm+$^{160}$Gd considering the disappearance of the capture pocket and the rapid increase of the potential with decreasing of the relative distance, it might produce new neutron-rich nuclide during the deep inelastic scattering process. Here, we study the isotope distribution of fragments in $^{154}$Sm+$^{160}$Gd from central to peripheral collisions with the ImQMD-v2.2 model \cite{Wang16}. Before investigating the isotope distribution of fragments in Sm+Gd, we have already tested the ImQMD-v2.2 model for description of the isotope distribution in the multi-nucleon transfer of $^{86}$Kr+$^{64}$Ni at an incident energy of 25 MeV/nucleon \cite{Sou02}. The measured isotope distribution of products can be reasonably well reproduced by using the ImQMD model together with a statistical code (GEMINI \cite{Char88}) for describing the secondary decay of fragments. We create 30000 events for each impact parameter and the ImQMD simulations are performed till $t=2000$ fm/c. Fig. 4 shows the predicted isotope distributions at $E_{\rm c.m.}=440$ MeV and at different impact parameters. The contour plots show the production probabilities of fragments in logarithmic scale. The curves denote the $\beta$-stability line described by Green's formula and the circles denote the nuclei with known masses in AME2012 \cite{Audi12}. At central collisions, broad charge and mass distributions of the products can be observed evidently due to large mass transfer between two nuclei. Some extremely neutron-rich fragments with $58 \le Z \le 76$ are observed in the ImQMD simulations. Through evaporating several neutrons from these fragments during the de-excitation, some new neutron-rich isotopes might be produced. With the increase of impact parameter, the number of transferred nucleons decrease due to the decrease of contact-times at peripheral collisions.

In Table 1, we list the production cross sections of some neutron-rich heavy nuclei with unknown masses. Here, we only list the nuclei with cross sections larger than $20 {\rm \mu b}$. Through multi-nucleon transfer in the deep inelastic scattering reaction of $^{154}$Sm+$^{160}$Gd, more than 40 neutron-rich nuclei with unknown masses can be produced, which implies that the deep inelastic scattering between two lanthanides is an efficient way to synthesize new neutron-rich heavy nuclei. The production cross sections decrease exponentially with further increasing of neutrons in an isotope chain. The production cross section of $^{182}$Yb is about $46 {\rm \mu b}$. In the table, the predicted mass excesses of these nuclei from a macroscopic-microscopic mass model, Weizs\"acker-Skyrme (WS4) model \cite{Wang14} are also presented.

\begin{figure}
\includegraphics[angle=0,width=0.8 \textwidth]{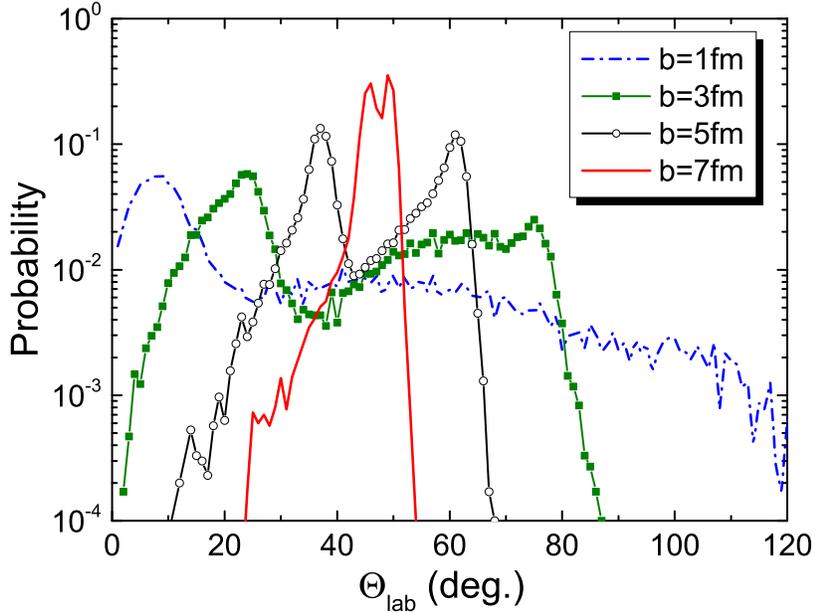}
\caption{(Color online) Angular distribution of heavy fragments with $Z\ge 62$ in $^{154}$Sm+$^{160}$Gd at $E_{\rm c.m.}=440$ MeV.}
\end{figure}

Simultaneously, the angular distribution of these neutron-rich nuclei produced in the deep inelastic scattering process are analyzed. Fig. 5 shows the calculated angular distribution of heavy fragments with $Z\ge 62$ from the ImQMD simulations. The different curves denote the results at different impact parameter. Most heavy fragments are emitted from forward angles. The double-peak structure of the angular distribution at the semi-central collisions can be clearly observed. We note that the sum of the two emission angles of the projectile-like and target-like nuclei is about $96^\circ$ (the corresponding value is $90^\circ$ in elastic scattering between two identical particles), which represents the behavior of elastic and inelastic scattering between the nearly-symmetric reaction partners. We also note that the neutron-rich fragments with unknown masses are mainly emitted from angles $\Theta_{\rm lab}<60^\circ$. Considering the relatively large production cross sections from the semi-central collisions (e.g, $b=5$ fm) and the angles of direct beams, $20^\circ <\Theta_{\rm lab}<60^\circ$ might be a suitable angular range to detect the neutron-rich heavy nuclei.

 \begin{table}
 \caption{ Production cross sections of some neutron-rich nuclei with unmeasured masses. The predicted mass excesses of these nuclei from the WS4 model \cite{Wang14} are also listed.}
\begin{tabular}{cccccccc }
\hline
$~~~Z~~~$ & $~~~N~~~ $ & $~~~\sigma ({\rm \mu b})~~~ $ & Mass Excess (MeV) & $~~~~~Z~~~$ & $~~~N~~~ $ & $~~~\sigma ({\rm \mu b})~~~ $ & Mass Excess (MeV) \\
\hline
58  &   94  &   65  & $-59.33$ &    68  &   105 &   975 & $-53.77$  \\
62  &   100 &   31  & $-54.52$ &    68  &   106 &   549 & $-52.31$  \\
63  &   101 &   256 & $-52.74$ &    68  &   107 &   96  & $-48.81$  \\
63  &   102 &   54  & $-50.36$ &    68  &   108 &   186 & $-46.89$  \\
64  &   100 &   1543& $-59.72$ &    68  &   109 &   100 & $-42.98$  \\
64  &   101 &   339 & $-56.29$ &    68  &   110 &   44  & $-40.63$  \\
64  &   102 &   194 & $-54.48$ &    69  &   108 &   316 & $-47.63$  \\
64  &   103 &   104 & $-50.62$ &    69  &   109 &   92  & $-44.28$  \\
65  &   100 &   3288& $-60.40$ &    69  &   110 &   35  & $-42.02$  \\
65  &   102 &   806 & $-55.82$ &    70  &   109 &   463 & $-46.71$  \\
65  &   103 &   140 & $-52.53$ &    70  &   110 &   186 & $-44.99$  \\
65  &   104 &   161 & $-50.30$ &    70  &   111 &   195 & $-41.38$  \\
65  &   105 &   46  & $-46.41$ &    70  &   112 &   46  & $-39.36$  \\
66  &   104 &   469 & $-53.98$ &    71  &   111 &   153 & $-41.77$  \\
66  &   105 &   130 & $-50.18$ &    71  &   113 &   59  & $-36.45$  \\
66  &   106 &   117 & $-47.99$ &    73  &   105 &   1639& $-50.32$  \\
67  &   105 &   656 & $-51.40$ &    73  &   116 &   138 & $-32.46$  \\
67  &   106 &   203 & $-49.33$ &    75  &   119 &   54  & $-27.29$  \\
67  &   107 &   145 & $-45.76$ &    76  &   121 &   180 & $-25.08$  \\

  \hline
\end{tabular}
\end{table}

\begin{center}
\textbf{V. SUMMARY}
\end{center}

In this work, we for the first time apply two different microscopic dynamics models for description of the deep inelastic scattering of $^{154}$Sm+$^{160}$Gd at energies above the Bass barrier. The fusion process is neither observed from the improved quantum molecular dynamics (ImQMD) simulations nor from the time dependent Hartree-Fock (TDHF) calculations.  The contact time of the di-nuclear system formed in head-on collisions is about 700 fm/c at an incident energy of 440 MeV, which is much shorter than the typical contact-times of quasi-fission. The time evolutions of the relative distance between the reaction partners at this energy from the two models are in good agreement with each other. Through investigating the nucleus-nucleus potential, we find that the capture pocket in $^{154}$Sm+$^{160}$Gd generally disappears, which leads to that the deep inelastic scattering process is a dominant process at central collisions. The isotope distribution of fragments in the deep inelastic scattering process is calculated with the ImQMD-v2.2 model together with the statistical decay model (GEMINI) for describing the secondary decay of fragments. More than 40 extremely neutron-rich nuclei with unknown masses are observed and the production cross sections are at the order of ${\rm \mu b}$ to mb. The multi-nucleon transfer in the deep inelastic scattering reaction of Sm+Gd seems to be an efficient way to produce new neutron-rich lanthanides. By analyzing the angular distribution of the produced heavy fragments, we suggest that $20^\circ <\Theta_{\rm lab}<60^\circ$ might be a suitable angular range to detect these extremely neutron-rich heavy nuclei.

\begin{center}
\textbf{ACKNOWLEDGEMENTS}
\end{center}
This work was supported by National Natural Science Foundation of
China (Nos. 11175252, 11275052, 11422548, 11547307, 11575189) and Guangxi Natural Science Foundation (No. 2015GXNSFDA139004). N. W. is grateful to Tong Wu and Hong Yao for performing the ImQMD calculations, and Zhong Liu for helpful discussions. L. G. acknowledge that the computations in this paper were performed in the Tianhe-1A supercomputer located in the Chinese National Supercomputer Center in Tianjin and the High-performance Computing Clusters of SKLTP/ITP-CAS.

\end{document}